# Comparison of Angular Spread for 6 and 60 GHz Based on 3GPP Standard

Jan M. Kelner, Cezary Ziółkowski, and Bogdan Uljasz
Institute of Telecommunications, Faculty of Electronics, Military University of Technology,
Warsaw, Poland
{jan.kelner, cezary.ziolkowski, bogdan.uljasz}@wat.edu.pl

*Abstract*—In an urban environment, a multipath propagation is one of the basic phenomena affecting a quality of received signals. This causes dispersions in time and angular domains. Basic parameters describing these dispersions are the rms delay spread and rms angle spread, respectively. The delay spread is related to a frequency of the transmitted signal and the nature of the propagation environment. In this paper, we show a mutual relationship between the time and angular dispersions in the received signal. The obtained simulation results present a comparison of the described dispersions for two different frequencies. In this case, the multi-elliptical propagation model and standard model developed by 3GPP are the basis for the simulation analysis of new communication system solutions.

## I. Introduction

Dynamic development of wireless communications began at the end of the 20th century with development of microelectronics and microwave technologies. At that time, a possibility of using microwave frequency ranges enabled creating new wireless communication systems operating in licensed and unlicensed bands. In this way, second (2G), third (3G), and then fourth generation (4G) mobile cellular systems and wireless local area networks (WLAN) based on IEEE 802.11 standards was created. At the beginning of the 21st century, the further development of microelectronics, nanoelectronics, and photonics contributed to using the higher frequency ranges from millimeter to terahertz waves. Currently, the millimeter waves are used, i.a., in the Internet of things (IoT) and emerging fifth generation (5G) communication systems.

Modern wireless systems are used primarily in urban areas or indoor environments. A characteristic feature of these environment types is multipath propagation. The negative impact of multipath propagation on quality of received signals is particularly evident in non-line-of-sight (NLOS) conditions. Dispersions of the received signal in time and angle domains is its effect.

Different lengths of propagation paths between a transmitter (Tx) and receiver (Rx) cause spreading an arrival time of the signal, and in practice, lead to intersymbol interference. A power delay profile (PDP) or spectrum (PDS) are instantaneous and averaged characteristics, respectively, that illustrate a time-dispersion level of radio channel. PDP/PDS are energetic characteristics based on the channel impulse response. Whereas, the rms delay spread (DS) is a measure of this dispersion. This parameter is often used in standard models for classification of the environment type, e.g., COST (*European Cooperation in Science and Technology*) 207 [1], WINNER (*Wireless World Initiative New Radio*) II [2], or 3GPP (*3rd Generation Partnership Project*) [3][4].

Whereas, different arrival directions of the signal components to the Rx is the cause of the angular dispersion in the received signal. A practical characteristic describing this phenomenon is the power angular spectrum (PAS), which is determined in the azimuth and elevation planes. Usually, the PAS is determined only in the azimuth plane and then, this acronym means the power azimuth spectrum. In contrast, the probability density function (PDF) of angle of arrival (AOA) is more often used in theoretical analyzes. Whereas, the rms angle spread (AS) is a measure of the angular dispersion.

New antenna techniques like multiple-input multiple-output (MIMO), massive MIMO [5][6], active phased array antenna (APAA), massive APAA [7], and beamforming [8] are the basis of the emerging 5G systems. In this case, knowledge of AOA statistic properties plays a significant role in the construction of antenna arrays and radio resource management in the 5G systems [9].

Empirical research and propagation models created on their basis show that DS decreases with increasing frequency for a given environment type. This is well illustrated in the 3GPP standard [4, Table 7.7.3-2]. On the other hand, AS and DS are strongly correlated linearly, what is shown in [10, Fig. 8][11, Fig. 1]. In this paper, we present comparison of AS for two frequencies of 6 and 60 GHz, which correspond to super high (SHF) and extremely high frequency (EHF) ranges, respectively. EHF is commonly called the millimeter wave range. The analysis is based on simulation studies. In this case, we use the multi-elliptical propagation model [12], which considers transmitting and receiving antenna patterns. PDP/PDS is the basic input data for this model. In the presented research, we use the PDP from the standard model developed by 3GPP [4].

The layout of the paper is as follows. Section II describes the PAS propagation model that considers antenna patterns.

The comparison of AS for two selected frequencies is presented in Section III. Conclusion is contained in Section IV.

## II. MULTI-ELLIPTICAL PROPAGATION MODEL

The genesis of multi-elliptical propagation models dates back to 1982, when Parsons and Bajwa presented the first approach of such model [13]. The premise for using confocal ellipses as locations of signal scatterers are empirical PDPs, in which characteristic time-clusters can be distinguished. In this case, a specific delay and power is assigned to each cluster. Thus, in the radio channel, the power accumulation for the specific delays indicates the grouping the propagation paths with the same length corresponding to the PDP delays. Hence, in the standard models, PDPs are usually presented in the form of discrete sets of the powers and delays. In the multi-elliptical models, the Tx and Rx are located in foci of the ellipses. This approach to geometric modeling gives the opportunity to faithfully reflect the empirical nature of propagation phenomena.

In the presented simulation studies, we use the multi-elliptical model, which additionally considers the local scattering occurring around the receiving antenna [14][15][16]. To describe this type of scattering, the von Mises distribution is used [17]. In [14]-[16], omnidirectional antennas for the Tx and Rx are assumed. However, a three-dimensional version of the model, which additionally considers the elevation plane, the transmitting and receiving antenna patterns is presented in [18] [19]. In this case, the scatterers are located on multi-halfelipsoid planes. An analogous method of modeling limited only to the azimuth plane is shown in [12]. This approach is used in this paper. Below, a brief description of it is presented.

Input data for the multi-elliptical propagation model are: PDP/PDS, Tx-Rx distance, number of paths for each time-cluster, parameters of the transmitting and receiving antennas, i.e., gain and half power beamwidth (HPBW) in the azimuth plane, for each antenna, respectively [12]. In addition, for the local scattering components, a parameter defining the von Mises distribution is specified. In the case of modeling line-of-sight (LOS) conditions, the Rice factor is additionally used to determine the contribution of the direct path component [14].

The number of clusters determines the number of ellipses, $N$, whereas, the sizes of the ellipses are defined by the PDP/PDS delays [14]. The transmitting antenna pattern is used to determine the PDF of angle of departure (AOD) [18]. In the case of omnidirectional or isotropic antennas, a uniform distribution is used [12]. For directional antennas, we use the Gaussian to model the patterns of these antennas [20]. Deviation of this distribution is related to the HPBW antenna. Then, the AODs are generated for each ellipse based on the determined PDF of AOD.

The AOA, $\varphi_R$, is determined for each AOD, $\varphi_T$, based on the following relationship [21]:

$$\varphi_R = \text{sgn}(\varphi_T) \arccos\left(\frac{2e_i + (1+e_i^2)\cos\varphi_T}{1+e_i^2 + 2e_i\cos\varphi_T}\right) \quad (1)$$

where $e_i$ is the eccentricity of the $i$th ellipse, $i=1,2,..,N$.

The powers of the individual time-clusters defined by PDP are the basis for determining uniform distributions. These distributions are used to generate the power of propagation paths that reach at the Rx surroundings from the individual ellipses [21]. Then, for directional antennas, the generated powers are modified by the receiving antenna pattern [19]. In this case, the Gaussian is also used.

As a result of the described algorithm, we obtain the sets of AOAs and powers. Based on these sets, PDF of AOA, $f(\varphi_R)$, is estimated [12]. In this paper, the so-defined PDF of AOA are used to assess the angular dispersion of the received signals.

## III. COMPARISION OF ANGLE SPREAD FOR 6 AND 60 GHz

The comparative analysis of the angular dispersion for two frequency bands, i.e., SHF and EHF, is carried out on the basis of simulation studies. In these studies, the multi-elliptical propagation model is used. On the one hand, this model allows considering the patterns of the transmitting and receiving antennas [12]. On the other hand, it allows accurate mapping the measurement data [16].

PDP is one of the basic input data of the multi-elliptical model. In simulation studies, we use a universal PDP for NLOS conditions described in the 3GPP standard [4, Table 7.7.2-2]. This PDP can be used for different frequency ranges, because the normalized delays are multiplied by DS, which characterizes a specific frequency and environment type [4, Table 7.7.3-2]. We use DSs corresponding to the so-called normal-delay profiles for the urban macro environment (UMa). In this case, for the analyzed frequencies, i.e., for 6 and 60 GHz, DSs are 363 and 228 ns, respectively. All simulations are obtained for the Tx-Rx distance of 200 m.

AS describing the angular dispersion and is defined as [20]

$$\sigma_\varphi = \sqrt{\int_{-180°}^{180°} \varphi_R^2 f(\varphi_R) \mathrm{d}\varphi_R - \left(\int_{-180°}^{180°} \varphi_R f(\varphi_R) \mathrm{d}\varphi_R\right)^2} \quad (2)$$

To evaluate the influence of the frequency range on the angular dispersion of the received signals, in simulation studies for two frequency ranges, we use antennas with the similar parameters. In the case of the millimeter waves, antennas are usually directional with a narrow beam. For example, in [22], the antenna HPBW used in measurements at 60 GHz is 7.3°. In the presented analysis, for each frequency, two types of horn antennas are included. For 60 GHz, the antenna parameters are as follows [23]: (PE9881-20) $G_A = 20$ dBi, $HPBW_A = 20°$, and (PE9881-24) $G_B = 24$ dBi, $HPBW_B = 12°$. For 6 GHz, we use similar antennas [24]: (ATH6G18) $G_D = 19$ dBi, $HPBW_D = 18°$, and (ATH800M6G) $G_D = 22$ dBi, $HPBW_D = 9°$.

In the analysis, we show the influence of the directions, $\alpha_T$ and $\alpha_R$, of the transmitting and receiving antennas, respectively, on AS for the analyzed antenna parameters. Figures 1 and 2 show the changes of AS versus $\alpha_T$ for $\alpha_R = 0°$ and the carrier frequency, $f_0$, equal to 60 or 6 GHz, respectively.

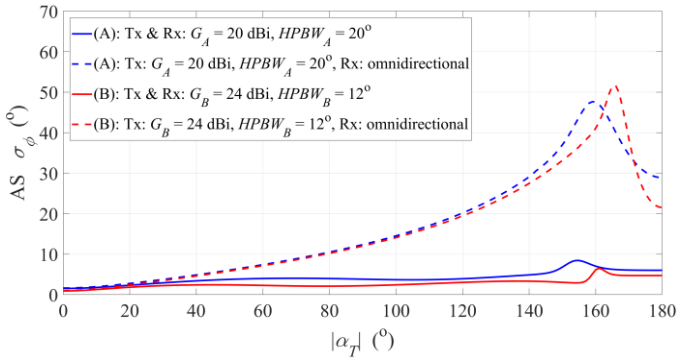

Fig. 1. AS versus $\alpha_T$ for $\alpha_R = 0°$ and $f_0 = 60$ GHz.

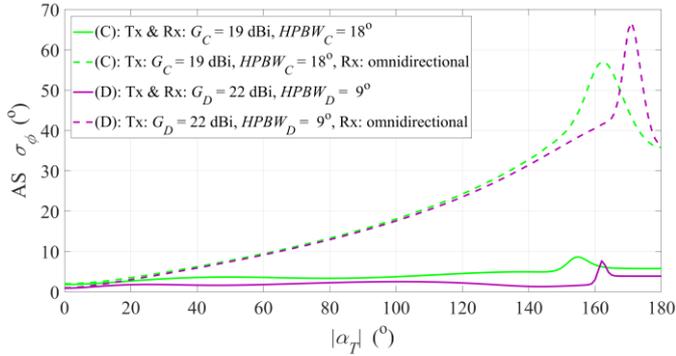

Fig. 2. AS versus $\alpha_T$ for $\alpha_R = 0°$ and $f_0 = 6$ GHz.

For each type of antenna, (A)–(D), we show two cases:

- the same antenna parameters are used for the transmitting and receiving antenna;
- the transmitting antenna is directional and the receiving antenna is omnidirectional.

In the second case, we can observe the effect of the angular dispersion around the receiving antenna. This effect allows to assess the influence of the transmitting antenna and channel on AS. Whereas, in the first case, we can see the effect of "spatial filtration" of scattering by the receiving antenna pattern. Then, we can observe that the selective directional antennas greatly reduce the angular dispersion of the environment.

The influence of the frequency on the angular dispersion is definitely more visible for the case with the omnidirectional receiving antenna. Then, AS is larger at 6 than at 60 GHz. This is in line with the conclusions in [10][11], where the increase of AS corresponds to the increase of DS.

The influence of the frequency, and thereby DS, on AS is practically invisible for two directional antennas. We can see better this effect in Fig. 3, where the respective graphs from Figs. 1 and 2 are shown. In this case, the antenna HPBW begins to play a significant role.

If the directions of the transmitting and receiving antennas are $\alpha_T = 0°$ and $\alpha_T = 0°$, respectively, then ASs are the smallest and amount to 1.5°, 0.9°, 1.8°, and 0.9° for (A)–(D), respectively. On the other hand, the total power of the received signal is usually the smallest for such orientation of the antennas.

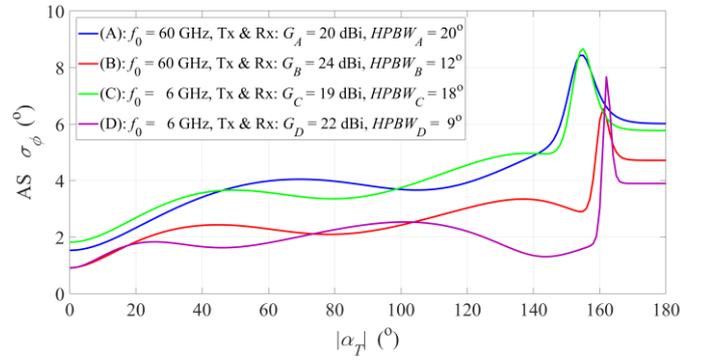

Fig. 3. Comparision AS versus $\alpha_T$ for $\alpha_R = 0°$, $f_0$ equal 60 GHz and 6 GHz.

The influence of $\alpha_R$ on AS for $\alpha_T = 180°$ is presented in Figs. 4 and 5, for 60 and 6 GHz, respectively.

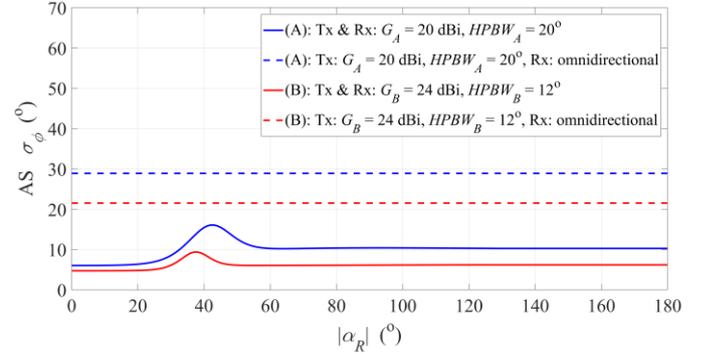

Fig. 4. AS versus $\alpha_R$ for $\alpha_T = 180°$ and $f_0 = 60$ GHz.

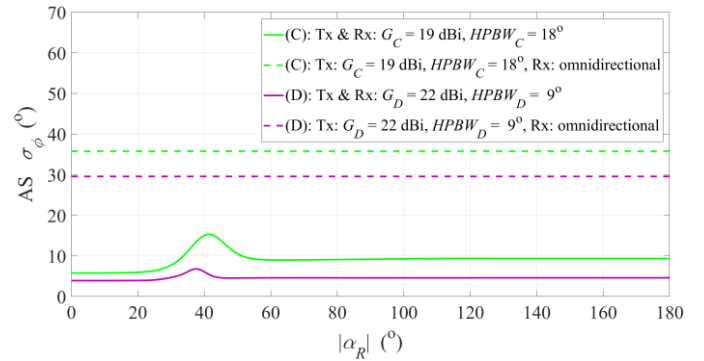

Fig. 5. AS versus $\alpha_R$ for $\alpha_T = 180°$ and $f_0 = 6$ GHz.

For the omni-directional receiving antenna and fixed $\alpha_T$, AS is fixed. Then, ASs are equal to 28.9°, 21.5°, 35.8°, and 29.6° for (A)–(D), respectively. In this case, we conclude like above that AS increases with frequency reduction.

If two directional antennas are included, the pattern of the receiving antenna also causes a significant reducing the angular dispersion. Figure 6 gives the possibility to compare all analyzed directional antenna type. In this case, for $\alpha_R = 0°$, ASs are the smallest and equal to 6.0°, 4.7°, 5.8°, and 3.9° for (A)–(D), respectively. For each of the analyzed antennas, a single extremum occurs in the AS graphs for $|\alpha_R| \approx 40°$. ASs for extremes are correlated with the antenna HPBWs. Values of $\alpha_R$ corresponding the maximum increases slightly with the increase of HPBW and basically does not depend on the frequency. For $|\alpha_R| > 60°$, ASs stabilize at a constant level and are amount to 10.3°, 6.2°, 9.3°, and 4.6° for (A)–(D), respectively. Thus, these values are equal to about half of the antenna HPBW.

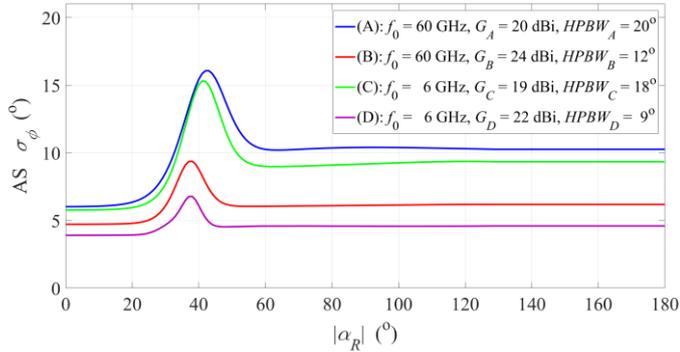

Fig. 6. Comparision AS versus $\alpha_R$ for $\alpha_T = 180°$, $f_0$ equal 60 GHz and 6 GHz.

For other orientations of the antenna (A) type, changes of AS versus $\alpha_T$ and $\alpha_R$ are shown in Figs. 7 and 8, respectively.

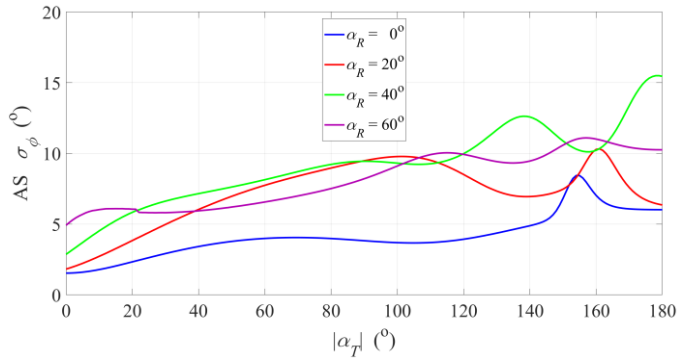

Fig. 7. AS versus $\alpha_T$ for antenna (A), $f_0 = 60$ GHz, and selected $\alpha_R$.

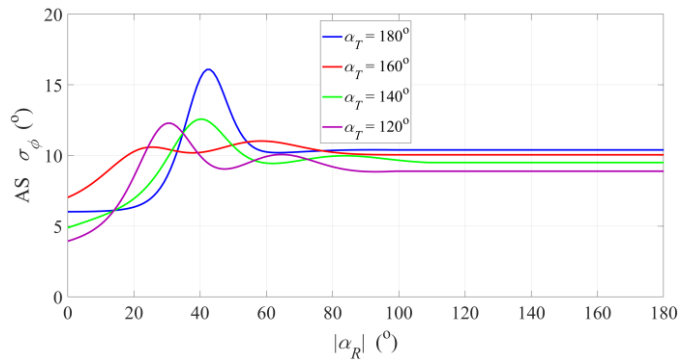

Fig. 8. AS versus $\alpha_R$ for antenna (A), $f_0 = 60$ GHz, and selected $\alpha_T$.

Based on the obtained graphs, the analysis of the angular dispersion is very difficult. However, we can draw general conclusions. On the one hand, the change of $\alpha_T$ has a smaller impact on the increase of AS than the change of $\alpha_R$. On the other hand, for a fixed $\alpha_T$, AS stabilizes above some $\alpha_R$, which is about 100° for $HPBW < 20°$.

## IV. CONCLUSION

New antenna techniques like MIMO, massive MIMO APAA, and beamforming are the basis of the emerging 5G systems. in this context, the knowledge of statistic properties of AOA plays a important role in the development of antenna arrays and radio resource management in the 5G systems. Most of available propagation models do not consider the directional antenna patterns. The proposed multi-elliptical model gives such opportunity. The carried out simulations show that the directional antennas significantly reduce the angular dispersion in the received signals. Therefore, including them to analyze modern communication systems is very important.

The purpose of the simulation studies was the comparative assessment of AS for 6 and 60 GHz. In these studies, the multi-elliptical propagation model, 3GPP standard and parameters of real antennas were used. Based on the obtained results, we could conclude that the influence of the frequency on the angular dispersion is visible only for the receiving antenna with the omnidirectional or sectoral wide-beam pattern. In the case of the narrow-beam directional antennas, the Rx antenna pattern ensures "spatial filtration" of the received signals. However, it should be remembered that simultaneously, this reduces the powers of the received signal components reaching from other directions. Additionally, the change of the receiving antenna direction affects the change of AS to a greater extent than the change of the transmitting antenna direction.

REFERENCES

[1] M. Failli, "COST 207. Digital land mobile radio communications," Commission of the European Communities, Directorate-General Telecommunications, Information Industries and Innovation, Luxembourg City, Luxembourg, Final report COST 207 (1984.03.14.-1988.09.13), 1989.

[2] "WINNER II channel models," IST-WINNER II, Tech. Rep. Deliverable 1.1.2 v.1.2., Sep. 2007.

[3] "3GPP TR 25.996 v13.1.0 (2016-12). Spatial channel model for multiple input multiple output (MIMO) simulations (Release 13)," 3rd Generation Partnership Project (3GPP), Technical Specification Group Radio Access Network, Valbonne, France, Tech. Rep. 3GPP TR 25.996 v13.1.0, Dec. 2016.

[4] "3GPP TR 38.901 V14.2.0 (2017-09). Study on channel model for frequencies from 0.5 to 100 GHz (Release 14)," 3rd Generation Partnership Project (3GPP), Technical Specification Group Radio Access Network, Valbonne, France, Tech. Rep. 3GPP TR 38.901 V14.2.0, Sep. 2017.

[5] R. Vannithamby and S. Talwar, Eds., *Towards 5G: Applications, requirements and candidate technologies*. Chichester, West Sussex, United Kingdom: Wiley, 2017.

[6] T. L. Marzetta, E. G. Larsson, H. Yang, and H. Q. Ngo, *Fundamentals of massive MIMO*. Cambridge, United Kingdom ; New York: Cambridge University Press, 2016.

[7] A. Taira *et al.*, "Performance evaluation of 44GHz band massive MIMO based on channel measurement," in *2015 IEEE Globecom Workshops (GC Wkshps)*, 2015, pp. 1–6.

[8] W. Liu and S. Weiss, *Wideband beamforming: Concepts and techniques*. Chichester, West Sussex, UK: Wiley, 2010.

[9] Y. Yang, J. Xu, G. Shi, and C.-X. Wang, *5G wireless systems: Simulation and evaluation techniques*. New York, NY, USA: Springer, 2017.

[10] K. I. Pedersen, P. E. Mogensen, and B. H. Fleury, "A stochastic model of the temporal and azimuthal dispersion seen at the base station in outdoor propagation environments," *IEEE Trans. Veh. Technol.*, vol. 49, no. 2, pp. 437–447, Mar. 2000.

[11] C. Ziółkowski and J. M. Kelner, "Empirical models of the azimuthal reception angle—Part II: Adaptation of the empirical models in analytical and simulation studies," *Wirel. Pers. Commun.*, vol. 91, no. 3, pp. 1285–1296, Dec. 2016.

[12] J. M. Kelner and C. Ziółkowski, "Modeling power angle spectrum and antenna pattern directions in multipath propagation environment," in *2018 12th European Conference on Antennas and Propagation (EUCAP)*, London, UK, 2018, pp. 1–5.

[13] J. D. Parsons and A. S. Bajwa, "Wideband characterisation of fading mobile radio channels," *IEE Proc. F Commun. Radar Signal Process.*, vol. 129, no. 2, pp. 95–101, Apr. 1982.

[14] C. Ziółkowski and J. M. Kelner, "Geometry-based statistical model for the temporal, spectral, and spatial characteristics of the land mobile channel," *Wirel. Pers. Commun.*, vol. 83, no. 1, pp. 631–652, Jul. 2015.

[15] C. Ziółkowski, "Statistical model of the angular power distribution for wireless multipath environments," *IET Microw. Antennas Propag.*, vol. 9, no. 3, pp. 281–289, Feb. 2015.

[16] C. Ziółkowski and J. M. Kelner, "Estimation of the reception angle distribution based on the power delay spectrum or profile," *Int. J. Antennas Propag.*, vol. 2015, p. e936406, Dec. 2015.

[17] A. Abdi, J. A. Barger, and M. Kaveh, "A parametric model for the distribution of the angle of arrival and the associated correlation function and power spectrum at the mobile station," *IEEE Trans. Veh. Technol.*, vol. 51, no. 3, pp. 425–434, May 2002.

[18] C. Ziółkowski and J. M. Kelner, "Antenna pattern in three-dimensional modelling of the arrival angle in simulation studies of wireless channels," *IET Microw. Antennas Propag.*, vol. 11, no. 6, pp. 898–906, May 2017.

[19] C. Ziółkowski and J. M. Kelner, "Statistical evaluation of the azimuth and elevation angles seen at the output of the receiving antenna," *IEEE Trans. Antennas Propag.*, vol. 66, 2018.

[20] R. Vaughan and J. Bach Andersen, *Channels, propagation and antennas for mobile communications*. London, UK: Institution of Engineering and Technology, 2003.

[21] C. Ziółkowski, J. M. Kelner, L. Nowosielski, and M. Wnuk, "Modeling the distribution of the arrival angle based on transmitter antenna pattern," in *2017 11th European Conference on Antennas and Propagation (EUCAP)*, Paris, France, 2017, pp. 1582–1586.

[22] T. S. Rappaport, G. R. MacCartney, M. K. Samimi, and S. Sun, "Wideband millimeter-wave propagation measurements and channel models for future wireless communication system design," *IEEE Trans. Commun.*, vol. 63, no. 9, pp. 3029–3056, Sep. 2015.

[23] "60 GHz WR-15 antennas," *Pasternack. The engineer's RF source*. [Online]. Available: https://www.pasternack.com/pages/Featured_Products/60-ghz-wr-15-antennas.html. [Accessed: 29-Jan-2018].

[24] "Microwave horns, antenna," *AR World. RF/microwave instrumentation*. [Online]. Available: https://www.arworld.us/html/13100.asp?S=3. [Accessed: 29-Jan-2018].